\documentclass[12pt]{article}
\usepackage{amscd,amssymb,amsmath,latexsym,enumerate}
\usepackage[mathscr]{euscript}
\usepackage{epsfig}
\usepackage{fancybox}
\usepackage{verbatim}
\usepackage{stmaryrd}

\usepackage[small,nohug]{diagrams}
\diagramstyle[labelstyle=\scriptstyle]

\usepackage{color}

\title{The spectral localizer for even index pairings}

\author{Terry A. Loring$^1$ and Hermann Schulz-Baldes$^2$
\\
\\
\\
{\small $^1$Department of Mathematics and Statistics, University of New Mexico, USA}
\\
{\small $^2$Department Mathematik, Friedrich-Alexander-Universit\"at Erlangen-N\"urnberg, Germany}
}

\date{ }

\newtheorem{theo}{Theorem}
\newtheorem{defini}{Definition}
\newtheorem{proposi}{Proposition}
\newtheorem{lemma}{Lemma}
\newtheorem{coro}{Corollary}

\newcommand{\CM}{{\mathbb C}}

\newcommand{\RM}{{\mathbb R}}
\newcommand{\SM}{{\mathbb S}}

\newcommand{\ZM}{{\mathbb Z}}

\newcommand{\DM}{{\mathbb D}}

\newcommand{\Aa}{{\cal A}}

\newcommand{\Bb}{{\cal B}}

\newcommand{\Ff}{{\cal F}}

\newcommand{\Qq}{{\cal Q}}
\newcommand{\Kk}{{\cal K}}
\newcommand{\Hh}{{\cal H}}

\newcommand{\one}{{\bf 1}}

\newcommand{\Ch}{{\rm Ch}} 
\newcommand{\Ind}{{\rm Ind}}

\newcommand{\Ran}{{\rm Ran}} 
 
\newcommand{\Sig}{{\rm Sig}} 
\newcommand{\diag}{{\rm diag}}

\newcommand{\SLoc}{L} 

\newcommand{\Tape}{F_\rho} 
 
\newcommand{\TapeOne}{F_1} 
\newcommand{\TapeD}{f_\rho}
\newcommand{\TapeDOne}{f_1}
\newcommand{\IndOne}{\psi} 
\newcommand{\IndTwo}{\phi} 
\newcommand{\IndOneD}{\Psi} 
\newcommand{\IndTwoD}{\Phi}

\begin{document}

\maketitle

\begin{abstract}
Even index pairings are integer-valued homotopy invariants combining an even Fredholm module with a $K_0$-class specified by a projection. Numerous classical examples are known from differential and non-commutative geometry and physics. Here it is shown how to construct a finite dimensional selfadjoint and invertible matrix, called the spectral localizer, such that half its signature is equal to the even index pairing. This makes the invariant numerically accessible. The index-theoretic proof heavily uses fuzzy spheres.
\end{abstract}

\section{Overview}

\subsection{Even Fredholm modules and index pairings}

Even index pairings involve a selfadjoint invertible  $H$ on a separable Hilbert space $\Hh$ which is paired with a so-called even  unbounded Fredholm module given by a Dirac operator $D$. A Fredholm module is also called a spectral triple or an unbounded $K$-cycle. From the pairing results a Fredholm operator and thus a Noether index. In the literature \cite{Con,GVF}, $H$ is supposed to lie in a given $C^*$-algebra $\Aa$, or a matrix algebra over $\Aa$, and then specifies a class in the $K_0$-group $K_0(\Aa)$ via the projection $P=\frac{1}{2}(\one-H|H|^{-1})$. The Fredholm module usually involves representations of $\Aa$. Here we  rather work with a hands-on purely operator theoretic approach in which $\Aa$ is simply the enveloping commutative algebra of $H$. 

\begin{defini}
\label{def-FredholmMod} An even Fredholm module for an invertible operator $H=H^*$ on $\Hh$ is a selfadjoint, invertible operator $D$ on $\Hh\oplus\Hh$ with compact resolvent and a selfadjoint unitary grading operator $\Gamma=\diag(\one,-\one)$ such that $\Gamma D\Gamma=-D$ and the commutator $[H\oplus H,D]$ extends to a bounded operator. Going into the eigenbasis of $\Gamma$, the operator $D$ decomposes
$$
D\;=\;
\begin{pmatrix} 0 & D_0^* \\ D_0  & 0 \end{pmatrix}
\;,
$$ 
with an invertible unbounded operator $D_0$ on $\Hh$. One then extracts a unitary operator $F$ on $\Hh$: 
$$
F\;=\;D_0\,|D_0|^{-1}
\;.
$$
The operator $D$ is called the Dirac operator. The identity $\Gamma D\Gamma=-D$ is also referred to as the chirality of $D$, and then $F$ is called the Dirac phase. 
\end{defini}

The following result is well-known, again {\it e.g.} \cite{Con} or p. 462 in \cite{GVF}.

\begin{theo}
\label{theo-pairings} 
An even Fredholm module for a selfadjoint invertible $H$ leads to a bounded Fredholm operator on $\Hh$
\begin{equation}
\label{eq-EvenPair}
T\;=\;P\,F\,P\,+\,(\one-P)
\;,
\end{equation}
where $P=\chi(H<0)$ is the spectral projection of $H$ on the negative spectrum and $F$ is the Dirac phase of the chiral Dirac operator.  
\end{theo}

\begin{defini}
\label{def-FredholmModInd} Given an even Fredholm module for an invertible $H=H^*$, the associated Fredholm operator $T$ and Noether index $\Ind(T)$ is referred to as the even index pairing. 
\end{defini}

The aim in the following is to provide a new approach to the calculation of $\Ind(T)$ which in concrete situations allows its evaluation by numerical computation. Indeed, it will be shown that it is given in terms of a finite dimensional matrix called the associated spectral localizer. In a prior paper \cite{LS} a similar result was obtained for pairings of odd Fredholm modules with $K_1$-classes. The construction in the even case is different and the proof is considerably more involved. In particular, a key step is a special deformation of the two-sphere (see Proposition~\ref{prop-SphereToSphere} below) that is similar to a map from the two-torus to the two-sphere used in prior works \cite{EL,HL}.

\subsection{Spectral localizer of an even index pairing}

The spectral localizer associated to an even Fredholm module $D$ for an invertible selfadjoint $H$ is by definition the operator
$$
\SLoc_{\kappa}
\;=\;
\begin{pmatrix} H & \kappa\,D_0^*\\ \kappa\,D_0 & - H \end{pmatrix}
\;=\;
\kappa\,D\,+\,H\otimes\Gamma
\;,
$$
acting on $\Hh\oplus\Hh$. Here $\kappa>0$ is a tuning parameter comparing $D$ to $H$ which we will assume to satisfy $\|H\|\geq 1$. We will consider finite volume restrictions of the spectral localizer w.r.t.\  the spatial structure given by the spectrum of the Dirac operator. Let $\pi_\rho$ be the surjective partial isometry onto the finite dimensional subspace $(\Hh\oplus\Hh)_\rho=\Ran\big(\chi(D^2\leq \rho^2)\big)$ of $\Hh\oplus\Hh$. Here $\chi$ is the characteristic function and $\rho>0$. For any operator $T$ on $\Hh\oplus\Hh$, let us then set $T_\rho=\pi_\rho T\pi_\rho^*$ which is simply the restriction of $T$ with Dirichlet boundary conditions. In particular, $\one_\rho=\pi_\rho\pi_\rho^*$ is the identity of $(\Hh\oplus\Hh)_\rho$. With these notations, the finite volume spectral localizer is the selfadjoint matrix acting on $(\Hh\oplus\Hh)_\rho$ given by
\begin{equation}
\label{eq-BottFinite}
\SLoc_{\kappa,\rho}
\;=\;
\big(\kappa\,D_\rho\,+\,H\otimes\Gamma\big)_\rho
\;.
\end{equation}
We will mainly be interested in the signature of $\SLoc_{\kappa,\rho}$.  Of course, one first has to assure that it is well-defined. This is the object of the following result, the proof of which is analogous to that of \cite[Theorem~5]{LS}.

\begin{theo}
\label{theo-invertible}
Let $g=\|H^{-1}\|^{-1}>0$ be the gap of $H$ and suppose that $\kappa$ and $\rho$ are such that
\begin{equation}
\label{eq-MainCond1}
\|[D,H\oplus H]\|\;\leq\;\frac{g^3}{12\,\|H\|\,\kappa}
\;.
\end{equation}
and
\begin{equation}
\label{eq-MainCond2}
\frac{2\,g}{\kappa}\;<\;\rho
\;.
\end{equation}
Then the matrix $\SLoc_{\kappa,\rho}$ is invertible and satisfies $(\SLoc_{\kappa,\rho})^2\geq\frac{g^2}{4}\,\one_\rho$. Moreover, the value of  $\SLoc_{\kappa,\rho}$ is independent of the choices of $\kappa$ and $\rho$, as long as \eqref{eq-MainCond1} and \eqref{eq-MainCond2} hold.
\end{theo}

Having a bounded commutator $[D,H\oplus H]$ is interpreted as a non-commutative differentiability of $H$ w.r.t.\ to the differential structure induced by $D$. If this is given, one can always choose the tuning parameter $\kappa$ sufficiently small so that \eqref{eq-MainCond1} holds, and in a second step the radius $\rho$ sufficiently large such that \eqref{eq-MainCond2} holds. It is also possible to revert the logic, namely first choose $\rho$ and then $\kappa$:

\begin{coro}
\label{coro-stable}
Let $g=\|H^{-1}\|^{-1}>0$ be the gap of $H$. If $\rho$ is such that
\begin{equation}
\label{eq-MainCond3}
\rho\;>\;\frac{24\,\|H\|\,\|[D,H\oplus H]\|}{g^2}
\;,
\end{equation}
and $\kappa$ such that
\begin{equation}
\label{eq-MainCond4}
\frac{2\,g}{\rho}\;<\;\kappa\;\leq\;
\frac{12\,\|H\|\,\|[D,H\oplus H]\|}{g^3}
\;,
\end{equation}
then $\SLoc_{\kappa,\rho}$ is invertible,  satisfies $(\SLoc_{\kappa,\rho})^2\geq\frac{g^2}{4}\,\one_\rho$, and $\Sig(\SLoc_{\kappa,\rho})$ is independent  of the choices of $\rho$ and $\kappa$ in this range. 
\end{coro}

We do not claim (or believe) that the conditions \eqref{eq-MainCond1}  and \eqref{eq-MainCond2} or \eqref{eq-MainCond3} and \eqref{eq-MainCond4} are optimal, but numerical results show that they cannot be improved by much. Clearly, $\SLoc_{\kappa,\rho}$ is invertible with vanishing signature when $\kappa=0$, and, more generally, when $\kappa$ is less than $g/\rho$. This shows that the first bound in \eqref{eq-MainCond4} is close to optimal. A similar argument shows any improved upper bound on $\kappa$ needs to be of order $O(1)$. As $\SLoc_{\kappa,\rho}$ is an even-dimensional selfadjoint and invertible matrix, its signature is indeed divisible by $2$.

\begin{defini}
\label{def-BottInd}
Whenever $\kappa$ and $\rho$ satisfy \eqref{eq-MainCond1} and \eqref{eq-MainCond2}, the integer $\frac{1}{2}\,\Sig(\SLoc_{\kappa,\rho})$ is called the localized index pairing of $H$ w.r.t.\ $D$. 
\end{defini}

Let us add a comment on the finite volume approximation. As $D^2=\diag(D_0^*D_0,D_0D_0^*)$, one has $(\Hh\oplus\Hh)_\rho= \Hh_{\rho,+}\oplus\Hh_{\rho,-}$ with  $\Hh_{\rho,+}=\Ran\big(\chi(D_0^*D_0 \leq \rho^2)\big) $ and $\Hh_{\rho,-}=\Ran\big(\chi(D_0D_0^* \leq \rho^2)\big)$.  Introducing the surjective partial isometries $\pi_{\rho,\pm}:\Hh\to\Hh_{\rho,\pm}$, one has $\pi_\rho=\pi_{\rho,+}\oplus\pi_{\rho,-}$ and 
$$
\SLoc_{\kappa,\rho}
\;=\;
\begin{pmatrix} \pi_{\rho,+} \,H\,\pi_{\rho,+} & \kappa\;\pi_{\rho,+}\,D_0^*\,\pi_{\rho,-}^* \\ \kappa\;\pi_{\rho,-}\,D_0\,\pi_{\rho,+}^* & - \pi_{\rho,-}\,H\,\pi_{\rho,-} \end{pmatrix}
\;.
$$
If $D_0$ is normal, then clearly $\Hh_{\rho,+}=\Hh_{\rho,-}$ and $\pi_{\rho,+}=\pi_{\rho,-}$. In the main result described next, we will make this simplifying assumption.

\subsection{Main result and comments}

The main result of this paper connects the Noether index of an even index pairing to the localized index pairing, namely the half-signature of the finite volume approximation $\SLoc_{\kappa,\rho}$ of the associated spectral localizer.  The main supplementary hypothesis is, from the perspective of non-commutative geometry, that the spacial coordinates encoded in the unbounded operator $D_0$ retains some commutativity. Theorem~\ref{theo-invertible} suggests that this may not be necessary, but several steps in the proof below would have to be considerably modified.

\begin{theo}
\label{theo-MainRes}
Let $D$ specify an even Fredholm module for an invertible selfadjoint $H$. Suppose that $\kappa$ and $\rho$ satisfy \eqref{eq-MainCond1} and \eqref{eq-MainCond2}. Furthermore assume that $D_0$ is normal.  Then the index pairing is given by: 
$$
\Ind(P\,F\,P\,+\,(\one-P))
\;=\;
\frac{1}{2}\,\Sig(\SLoc_{\kappa,\rho})
\;.
$$
\end{theo}

If the operator $H$ has some symmetry property involving a real structure, it may be possible to extract a $\ZM_2$-invariant as the sign of the determinant or Pfaffian, see \cite{Lor,LS}, and connect them to $\ZM_2$-indices as defined in \cite{GS,BKR}. These issues will be examined in a subsequent paper.

\vspace{.2cm}

Our initial motivation to prove this theorem rooted in applications to the field of topological insulators. In a one-particle and tight-binding approximation, these quantum systems  are described by a Hamiltonian $H$ on $\ell^2(\ZM^d,\CM^N)$ with spectral gap at the Fermi level  (making the system into an insulator), but also a topologically non-trivial Fermi projection $P=\chi(H<0)$ in the sense that it has non-trivial winding numbers or Chern numbers. An overview of the physics and mathematics literature on the subject is contained in \cite{HL,Lor,PSB}. Theorem~\ref{theo-MainRes} is of interest for systems in even spatial dimension and provides a very efficient means to numerically calculate the topological invariant. Initial numerical results are contained in \cite{Lor} and \cite{FPL}, but the method will be further explored elsewhere. For odd dimensional physical systems (with chiral symmetry), one has to consider odd index pairings of a invertible $A$ specifying a $K_1$-class with an odd Fredholm module given by a Dirac operator $D$ without chiral symmetry, see \cite{PSB}. If $\Pi=\chi(D<0)$ is the associated Hardy projection, one has a Fredholm operator $\Pi A\Pi+(\one-\Pi)$ with an index which can also be calculated as the signature of a suitably defined spectral localizer \cite{LS}. Before turning to the proof of Theorem~\ref{theo-MainRes}, let us  spell out in the next section how the even Fredholm module for a Hamiltonian describing a topological insulator is constructed and how the index pairing is related to Chern numbers.

\subsection{Examples from physics}

Let $d$ be even. The Hamiltonian $H$ is a selfadjoint and invertible, bounded operator on the Hilbert space $\Hh=\ell^2(\ZM^d,\CM^N)$. Furthermore, on $\Hh$ act  the $d$ components $X_1,\ldots,X_d$ of the selfadjoint commuting position operators defined by $X_j|n\rangle=n_j|n\rangle$ where $n=(n_1,\ldots,n_d)\in\ZM^d$ and $|n\rangle\in\ell^2(\ZM^d)$ is the Dirac Bra-Ket notation for the unit vector localized at $n$. Moreover, let be given a self-adjoint irreducible representation $\gamma_1,\ldots,\gamma_{d-1}$ of the Clifford algebra $\CM_{d-1}$ on $\CM^{N}$. Hence $N=2^{\frac{d}{2}}$. From this data, one sets 
$$
D_1
\;=\;
\sum_{j=1}^{d-1}X_j\otimes\gamma_j
\;+\;
|0\rangle\langle 0|\otimes \gamma_1
\;,
\qquad
D_2\;=\;X_d\otimes\one
\;.
$$
If then $\sigma_1=\binom{0\;1}{1\;0},\,\sigma_2=\binom{0\;-\imath}{\imath\;\;\;0},\,\sigma_3=\binom{1\;\;\;0}{0\;-1}$ are the standard Pauli matrices, the selfadjoint Dirac operator  is
\begin{equation}
\label{eq-StandDirac}
D\;=\;
D_1\otimes\sigma_1\;+\;
D_2\otimes\sigma_2
\;,
\end{equation}
or equivalently $D_0=D_1+\imath\,D_2$. Then $D$ is odd w.r.t.\ $\Gamma=\one\otimes\sigma_3$, namely $\Gamma D\Gamma=-\,D$.  Clearly, $D$ has a compact resolvent and $[D_1,D_2]=0$, namely the off-diagonal entry $D_0$ of $D$ is normal. It defines a Fredholm module for $H$ if 
\begin{equation}
\label{eq-LocalityBound}
\|[D_1,H]\|\,<\,\infty
\;,
\qquad
\|[D_2,H]\|\,<\,\infty
\;.
\end{equation}
From a physicists perspective, the bounds \eqref{eq-LocalityBound} express the locality of the Hamiltonian $H$, while for a mathematician it is rather the non-commutative differentiability of $H$.  The index pairing is known to be connected to the $d$-th Chern number by an index theorem, see \cite{PLB} and \cite[Corollary 6.3.2]{PSB}:
$$
\Ind(P\,F\,P\,+\,(\one-P))
\;=\;
\Ch_d(P)
\;.
$$
For the Dirac operator \eqref{eq-StandDirac}, one has $\Hh_{\rho,\pm}=\ell^2(\DM_\rho,\CM^N)$ where $\DM_\rho=\{x\in\ZM^d\,:\,\|x\|\leq \rho\}$ denotes the discrete disc of radius $\rho$. Therefore, the spectral localizer is really an operator localized in physical space.

\vspace{.2cm}

Most numerical studies of topological insulator have been conducted on square samples.  The truncation of the spectral localizer $L_\kappa$ can be made to a square $\{x\in\ZM^d\,:\,|x_j|\leq \rho\}$.  The proof of Theorem~\ref{theo-invertible} can be modified to show the signature is still equal to index when a square sample is used, or any sample that includes the disk of radius $\rho$.  One can also vary the local geometry of model, replacing $ \ZM^d$ by the vertices of a quasi-lattice \cite{FPL} or use more random collections of points, as in an amorphous system  \cite{BP}.  The spectral localizer method works in these cases since it uses Dirichlet boundary conditions.

\subsection{Outline of the proof}

The main technical tools in the proofs are fuzzy spheres. While they can be defined in arbitrary dimension, we will only work of fuzzy $2$-spheres an refer to them simply as fuzzy spheres. 

\begin{defini}
\label{def-FuzzySphere}
Let $\Kk$ be a C$^*$-algebra with unitization $\Kk^+$. A fuzzy sphere $(X_1,X_2,X_3)$ of width $\delta<1$ in $\Kk$ is a collection of three self-adjoints $X_1,X_2,X_3\in\Kk^+$ with spectrum in $[-1,1]$ such that, for $i,j=1,2,3$,
$$
\Big\|\one -(X_1^2+X_2^2+X_3^2)\Big\|\;<\;\delta\;,
\qquad
\|[X_j,X_i]\|\;<\;\delta
\;.
$$
\end{defini}

There is a tight link between fuzzy spheres and classes in the $K_0$-group $K_0(\Kk)$ of $\Kk$. While the standard description of this group is in terms of homotopy equivalence classes of projections in matrix algebras $M_n(\Kk)$ over $\Kk$, it is also possible ({\it e.g.} \cite{GS,LS}) to use homotopy equivalence classes $[L]_0$ of invertible selfadjoint matrices $L\in M_{2n}(\Kk^+)$ having a scalar part $s(L)\in M_{2n}(\CM)$ that is homotopic to $\diag(\one_n,-\one_n)$. The additive structure is then simply given by direct sum.

\begin{proposi}
\label{eq-SphereK0}
A fuzzy sphere $(X_1,X_2,X_3) $ of width $\delta\leq \frac{1}{4}$ in $\Kk$ specifies a class $[L]_0\in K_0(\Kk)$ by the self-adjoint invertible operator
$$
L
\;=\;
\sum_{j=1,2,3} X_j\otimes\sigma_j
\;\in\;M_2(\Kk^+)
\;,
$$
where $\sigma_1,\sigma_2,\sigma_3$ are the Pauli matrices and $M_2(\Kk^+)$ the $2\times 2$ matrices with entries in $\Kk^+$. 
\end{proposi}

\noindent {\bf Proof.} Indeed,
\begin{equation}
\label{eq-LSquare}
L^2
\;=\;
(X_1^2\,+\,X_2^2\,+\,X_3^2)\otimes\one\,+\sum_{i<j}\,[X_i,X_j]\otimes\sigma_i\sigma_j
\;,
\end{equation}
so that
$$
L^2\;\geq\;\one\,-\,\|X_1^2\,+\,X_2^2\,+\,X_3^2-\one\|\,-\,3\,\sup_{i<j}\|[X_i,X_j]\|
\;>\;(1\,-\,4\,\delta)\one
\;.
$$
Thus $\delta\leq \frac{1}{4}$ implies the invertibility of $L$.
\hfill $\Box$

\vspace{.2cm}

As the notation already suggests, we will show in Section~\ref{sec-SLocFuzzy} that the spectral localizer can be deformed into a fuzzy sphere within the space of compact operators, actually even inside the matrices of fixed size. More precisely, the following will be proved:

\begin{proposi}
\label{prop-SLocToSphere}
Suppose that $\kappa$ is chosen sufficiently small and $\rho$ sufficiently large so that, in particular,  \eqref{eq-MainCond1} and \eqref{eq-MainCond2} hold. Then there is a fuzzy sphere $(X_1,X_2,X_3)$ of matrices of half the size of $\SLoc_{\kappa,\rho}$ such that $\sum_{j=1,2,3} X_j\otimes\sigma_j$ is homotopic to $\SLoc_{\kappa,\rho}$ within the invertible self-adjoint matrices. The width of the fuzzy sphere $(X_1,X_2,X_3)$ is of the order $\rho^{-1}$.
\end{proposi}

Hence the spectral localizer can be identified with a fuzzy sphere. Actually, this fuzzy sphere will be constructed explicitly in Section~\ref{sec-SLocFuzzy}. On the other hand, also the index pairing \eqref{eq-EvenPair} itself can be identified with a fuzzy sphere. Indeed, it specifies a class $[\pi(T)]_1$ in the $K_1$-group $K_1(\Qq)$ of the Calkin algebra $\Qq$ over the Hilbert space $\Hh$ (here $\pi:\Bb\to\Qq$ denotes the projection from the bounded operators on $\Hh$ onto the Calkin algebra). Via the index map this class is mapped to an element in $K_0(\Kk)$ which under suitable conditions is given by a fuzzy sphere in $\Kk$. This is a consequence of the following abstract result.

\begin{theo}
\label{theo-IndSphere}
Let $0 \rightarrow \Kk \,\hookrightarrow\,\Bb\,\overset{\pi}{\rightarrow}\, \Qq \rightarrow 0$ be a short exact sequence of C$^*$-algebras with $\Qq$ unital. Let $A\in\Bb$ be a contraction such that $\pi(A)\in\Qq$ is invertible specifying an element $[\pi(A)]_1\in K_1(\Qq)$.  Set $A_1=\frac{1}{2}(A+A^*)\in\Bb$ and $A_2=\frac{1}{2\imath}(A-A^*)\in\Bb$ and assume that
\begin{equation}
\label{eq-LiftCommEst}
\| [A_1,A_2]\|
\;<\;\epsilon
\;,
\end{equation}
for some $\epsilon$ sufficiently small. Also introduce a non-negative operator $B=(A_1^2+A_2^2)^{\frac{1}{2}}$. Further let $\IndOne:[0,1]\to [0,1]$ and $\IndTwo:[0,1]\to [-1,1]$  be smooth functions with bounded derivatives such that
\begin{equation}
\label{eq-IndFunct}
\IndTwo(1)\,=\,1\,=\,-\IndTwo(0)
\;,
\qquad
x^2\,\IndOne(x)^4\,+\,\IndTwo(x)^2\,=\,1\;.
\end{equation}
Next set
\begin{align}
& Y_1
\;=\;
\IndOne(B)A_1\IndOne(B)
\;,
\nonumber
\\
&
Y_2
\;=\;
-\,\IndOne(B)A_2\IndOne(B)
\;,
\label{eq-YSphere}
\\
&
Y_3\;=\;\IndTwo(B)
\;.
\nonumber
\end{align}
Then the $K$-theoretic index map satisfies
$$
\Ind [\pi(A)]_1
\;=\;
\Big[
\sum_{j=1,2,3} Y_j\otimes\sigma_j
\Big]_0
\;,
$$
and $(Y_1,Y_2,Y_3)$ is a fuzzy sphere in $\Kk$ of width depending on $\epsilon$ as well as $\IndOne$ and $\IndTwo$.
\end{theo}

\noindent {\bf Proof.} It is well-known ({\it e.g.} Proposition~3 in \cite{LS}) that
$$
\Ind[\pi(A)]_1
\;=\;
\left[
\begin{pmatrix}
2 A^*A-\one & 2(\one-AA^*)^{\frac{1}{4}}A(\one-A^*A)^{\frac{1}{4}}
\\
2(\one-A^*A)^{\frac{1}{4}}A^*(\one-AA^*)^{\frac{1}{4}} & -(2AA^*-\one)
\end{pmatrix}
\right]_0
\;.
$$
As both $A^*A$ and $AA^*$ are close to $B^2$, one can replace without changing the $K_0$-class:
$$
\Ind[\pi(A)]_1
\;=\;
\left[
\begin{pmatrix}
2 B^2-\one & 2(\one-B^2)^{\frac{1}{4}}A(\one-B^2)^{\frac{1}{4}}
\\
2(\one-B^2)^{\frac{1}{4}}A^*(\one-B^2)^{\frac{1}{4}} & -(2B^2-\one)
\end{pmatrix}
\right]_0
\;.
$$
This shows that the result holds for $\IndOne(x)=\sqrt{2}(1-x^2)^{\frac{1}{4}}$ and $\IndTwo(x)=2x^2-1$. Of course, these functions can be homotopically changed without changing the class in $K_0(\Kk)$.  Choosing $\IndOne$ and $\IndTwo$ smooth, the assumption \eqref{eq-LiftCommEst} combined with smooth spectral calculus assures for the commutators 
$$
\|[A,\IndOne(B)]\|\;\leq\;C\,\epsilon
\;,
\qquad
\|[A,\IndTwo(B)]\|\;\leq\;C\,\epsilon
\;.
$$ 
This implies that $[Y_1,Y_3]=\IndOne(B)[A,\IndTwo(B)]\IndOne(B)$ is small in norm, and similarly $[Y_2,Y_3]$ and $[Y_1,Y_2]$ are also small. Furthermore, the last equation in \eqref{eq-IndFunct} assures $Y_1^2+Y_2^2+Y_3^2-\one$ is small, so that $(Y_1,Y_2,Y_3)$ indeed forms a fuzzy sphere.
\hfill $\Box$

\vspace{.2cm}

In the application of Theorem~\ref{theo-IndSphere} to the proof of Theorem~\ref{theo-MainRes}, the algebras $\Kk$ and $\Bb$ are respectively the compact and bounded operators on $\Hh$ and $\Qq$ is the Calkin algebra. The suitable lift $A$ with $\pi(A)=\pi(T)$ with $T$ given by \eqref{eq-EvenPair} is explicitly constructed in Section~\ref{sec-IndMap}. Let us note that by the lifting result in \cite[Theorem~3.1]{SL} one can always achieve \eqref{eq-LiftCommEst} if $\pi(A)$ is unitary. 

\vspace{.2cm}

Getting back to the main argument, we note that there is one fuzzy sphere $(X_1,X_2,X_3)$ in the algebra of compact operators $\Kk$ associated to the spectral localizer by Proposition~\ref{prop-SLocToSphere}, and another fuzzy sphere $(Y_1,Y_2,Y_3)$ in $\Kk$ associated to the index pairing  \eqref{eq-EvenPair} by Theorem~\ref{theo-IndSphere}. The proof of Theorem~\ref{theo-MainRes}, the main result of the paper, will then be completed by showing that these two fuzzy spheres can be deformed into each other within the space of fuzzy spheres. For this purpose, we will use a particular smooth map $\Ff:\SM^2\to\SM^2$ of mapping degree $1$.

\begin{proposi}
\label{prop-SphereToSphere}
Let $\IndOneD:[-1,1]\to [0,1]$ and $\IndTwoD:[-1,1]\to [-1,1]$  be continuous functions such that for $x\geq 0$:
\begin{equation}
\label{eq-IndFunctD}
\IndOneD(-x)=0\;,
\qquad
\IndTwoD(-x)=\IndTwoD(x)\;,
\qquad
\IndTwoD(0)\,=\,1\,=\,-\IndTwoD(1)
\;,
\qquad
(1-x^2)\,\IndOneD(x)^4\,+\,\IndTwoD(x)^2\,=\,1\;.
\end{equation}
For $(x_1,x_2,x_3)\in\SM^2$, set
\begin{align}
& 
z_1\;=\;\IndOneD(x_3)x_1\IndOneD(x_3)\,+\,\IndOneD(-x_3)(1-x_3^2)^{\frac{1}{2}}\IndOneD(-x_3)\;,
\nonumber
\\
& 
z_2\;=\;-\,\IndOneD(x_3)x_2\IndOneD(x_3)\;,
\label{eq-XZDef}
\\
&
z_3\;=\;\IndTwoD(x_3)
\;.
\nonumber
\end{align}
Then $\Ff(x_1,x_2,x_3)=(z_1,z_2,z_3)$ defines a continuous map $\Ff:\SM^2\to\SM^2$ which has mapping degree $1$. If $\IndOneD$ and $\IndTwoD$ are differentiable with bounded derivatives, there exists an explicit homotopy $\lambda\in[0,1]\mapsto \Ff_\lambda$ of differentiable degree $1$ maps of $\SM^2$ connecting $\Ff_1=\Ff$ to the identity $\Ff_0$. 
\end{proposi}

The proof of Proposition~\ref{prop-SphereToSphere} will be given in Section~\ref{sec-DeformFuzzy}. Of course, the above functions commute so that one can also write $z_2=\IndOneD(x_3)^2x_2$, and so on. However, when $\Ff$ is applied to a fuzzy sphere, the order of the factors is relevant. From the fuzzy sphere $(X_1,X_2,X_3)$ as given in Proposition~\ref{prop-SLocToSphere}, let us thus set
\begin{align}
& Z_1\;=\;\IndOneD(X_3)X_1\IndOneD(X_3)\,+\,\IndOneD(-X_3)(\one-X_3^2)^{\frac{1}{2}}\IndOneD(-X_3)\;,
\nonumber
\\
& Z_2\;=\;-\,\IndOneD(X_3)X_2\IndOneD(X_3)\;,
\label{eq-MappedFuzzy}
\\
& Z_3\;=\;\IndTwoD(X_3)
\;.
\nonumber
\end{align}
By construction, these operators are selfadjoint with spectrum in $[-1,1]$. It will also be verified in Section~\ref{sec-DeformFuzzy} that $(Z_1,Z_2,Z_3)$ forms a fuzzy sphere in $\Kk$ if $(X_1,X_2,X_3)$ does so. As $\Ff$ has degree $1$, it can be shown that the two fuzzy spheres $(X_1,X_2,X_3)$ and $(Z_1,Z_2,Z_3)$ lie in the same class in $K_0(\Kk)$ via Proposition~\ref{eq-SphereK0}. In Section~\ref{sec-CompFuzzy}  the proof is then concluded by showing that $(Y_1,Y_2,Y_3)$ defines the same $K_0$-class as $(Z_1,Z_2,Z_3)$, provided that the functions $\IndOne$, $\IndTwo$  and $\IndOneD$, $\IndTwoD$ are chosen dual to each other via the relations
\begin{equation}
\label{eq-duality}
\IndOne(x)\,=\,\IndOneD(\sqrt{1-x^2})\;,
\qquad
\IndTwo(x)\,=\,\IndTwoD(\sqrt{1-x^2})\;.
\end{equation}
Otherwise stated:

\begin{theo}
\label{theo-MainResInd}
Under the same assumptions as in {\rm Theorem~\ref{theo-MainRes}},
$$ 
\Ind\,[\pi(P\,F\,P\,+\one-P)]_1
\;=\;
[L_{\kappa,\rho}]_0
\;.
$$
\end{theo}

As $K_1(\Qq)\cong\ZM$ and $K_0(\Kk)\cong\ZM$, Theorem~\ref{theo-MainRes} follows immediately. This concludes the overview of the proof of Theorem~\ref{theo-MainRes}. The remainder of the paper provides the details.

\section{Technical tools}
\label{sec-Tools}

In the proofs below, $\kappa$ will always be chosen sufficiently small so that \eqref{eq-MainCond1} holds. Then it is kept fixed while $\rho$ is taken sufficiently large, often much larger then necessary for \eqref{eq-MainCond2}. The errors are then estimated in terms of $\rho^{-1}$. Therefore it is convenient to use the following notation. Given two nets $(A_\rho)$ and $(B_\rho)$ of bounded operators and with positive family index $\rho>0$ and given any $\gamma\geq 1$, we write $A_\rho\sim_{\rho^\gamma} B_\rho$ whenever there is a constant $C$ such that $\|A_\rho-B_\rho\|\leq C\rho^{-\gamma}$. By abuse of notation, we also simply write $A\sim_\rho B$ if it is clear from the context that $A=(A_\rho)$ and $B=(B_\rho)$ are nets indexed by $\rho$. We will use the following basic

\begin{lemma}
\label{lem-RhoBehave}
Let $A=(A_\rho)$ and $B=(B_\rho)$ be selfadjoint and uniformly bounded in $\rho$. Furthermore let us suppose that $A\sim_\rho B$. 
\begin{itemize}
\item[{\rm (i)}] If $f:\RM\to\RM$ is smooth on the spectrum of both $A$ and $B$, then $f(A)\sim_\rho f(B)$. 

\item[{\rm (ii)}] If, moreover, $P=P^2=P^*$ is an orthogonal projection, then $f(PAP)\sim_\rho Pf(A)P$.

\item[{\rm (iii)}] If $A$ and $B$ are non-negative and $\gamma\geq 1$, then $A^{\frac{1}{\gamma}}\sim_{\rho^{\gamma}} B^{\frac{1}{\gamma}}$.

\end{itemize}

\end{lemma}

\noindent {\bf Proof.} The claim (i) can be proved using from Dykin's functional calculus (often also called Helffer-Sjorstrand formula) 
$$
f(A_\rho)
\;=\;
\int_{\RM^2} \frac{dz}{2\pi}\;\partial_{\overline{z}}\widetilde{f}(z)\,(A_\rho-z)^{-1}
\;.
$$
Here $\widetilde{f}:\CM\to\CM$ is a suitable quasianalytic extension of $f$. Now invoking the resolvent identity 
$$
f(A_\rho)-f(B_\rho)
\;=\;
\int_{\RM^2} \frac{dz}{2\pi\imath}\;\widetilde{f}(z)\,(B_\rho-z)^{-1}(A_\rho-B_\rho)(A_\rho-z)^{-1}
\;.
$$
As $\widetilde{f}$ can be chosen with arbitrary decay on the real axis, the bound $\|A_\rho-B_\rho\|\leq C\rho^{-\gamma}$ implies 
$\|f(A_\rho)-f(B_\rho)\|\leq C\rho^{-\gamma}$ for some different constant $C$. For (ii) one can proceed in the same manner by using the geometric resolvent identity. Item (iii) directly follows from the bound $\|A_\rho^{\frac{1}{\gamma}}- B_\rho^{\frac{1}{\gamma}}\|\leq \|A_\rho-B_\rho\|^{\frac{1}{\gamma}}$, {\it e.g.} \cite[Lemma~1]{LS}.
\hfill $\Box$

\vspace{.2cm}

In the following sections it will be crucial to control the functional calculus of $D$ and $H$ with slow-varying functions. 

\begin{lemma}
\label{lem-Tapering}
Let $G_\rho:\RM\to\RM$ be a differentiable function of the form $G_\rho(x)=G_1(\frac{x}{\rho})$ with a derivative having an integrable Fourier transform $\|\widehat{G_1'}\|_{L^1(\RM)}<\infty$. Here the Fourier transform is defined by $\int dx\,e^{-\imath px} G'_1(x)$, so without a factor $2\pi$. For any bounded operator $A=A^*$ on $\Hh\oplus\Hh$, one then has
\begin{equation}
\label{eq-TapeCommuGen}
\|[G_\rho(D),A]\|
\;\leq\;
\rho^{-1}\,\|\widehat{G_1'}\|_{L^1(\RM)}\,\|[D,A]\|
\;,
\end{equation}
and
\begin{equation}
\label{eq-TapeCommuGen2}
\|[D,G_\rho(A)]\|
\;\leq\;
(2\pi\rho)^{-1}\,\|\widehat{G_1'}\|_{L^1(\RM)}\,\|[D,A]\|
\;.
\end{equation}
\end{lemma}

\noindent {\bf Proof.}
A short calculation shows that $\|\widehat{G'_\rho}\|_{L^1(\RM)}=\rho^{-1}\|\widehat{G_1'}\|_{L^1(\RM)}$. Therefore the first claim follows immediately from \cite[Theorem 3.2.32]{BR} or \cite[Lemma~10.15]{GVF}, and the second from \cite[Proposition~3.3.6]{Sak}. 
\hfill $\Box$

\vspace{.2cm}

We will choose a particular so-called tapering function $\TapeOne:\RM\to[0,1]$ which is even, vanishes on $\RM\setminus[-1,1]$ and is equal to $1$ on $[-\frac{1}{2},\frac{1}{2}]$. Lemma~4 of \cite{LS} explicitly constructs such a function with $\|\widehat{\TapeOne'}\|_{L^1(\RM)}\leq 8$. As in Lemma~\ref{lem-Tapering}, let us then set $\Tape(x)=\TapeOne(\frac{x}{\rho})$. The function $\Tape$ is supported by $[-\rho,\rho]$ and is equal to $1$ on $[-\frac{\rho}{2},\frac{\rho}{2}]$. The bound \eqref{eq-TapeCommu} holds for $G_\rho=\Tape$ and $A=H\otimes\sigma_0$ where $\sigma_0$ denotes the $2\times 2$ identity matrix:
\begin{equation}
\label{eq-TapeCommu}
\|[F_\rho(D),H\otimes\sigma_0 ]\|
\;\leq\;
8\,\rho^{-1}\,\|[D,H\otimes\sigma_0 ]\|
\;.
\end{equation}

\section{Invertibility of the spectral localizer}
\label{sec-InvSLoc}

The object of this section is to prove Theorem~\ref{theo-invertible}  and thus that the signature of the finite volume spectral localizer is well-defined and stable. Several elements of this proof will be used in Section~\ref{sec-SLocFuzzy}.

\vspace{.2cm}

\noindent {\bf Proof} of Theorem~\ref{theo-invertible}.  To connect different values of $\rho$, let us consider the matrix
$$
L_{\kappa,\rho,\rho'}(\lambda)
\;=\;
\kappa\;\pi_{\rho'}\,D\,\pi_{\rho'}^{*}
\;+\;
\pi_{\rho'}F_{\lambda,\rho}(H\otimes\Gamma)\, F_{\lambda,\rho}\,\pi_{\rho'}^{*}
\;,
$$
acting on $(\Hh\oplus\Hh)_{\rho'}$ where
$$
F_{\lambda,\rho} \;=\; (1-\lambda)\one\;+\;\lambda F_{\rho}(D)
\;,
$$
and $\rho\leq\rho'$ and $0\leq\lambda\leq1$, and \eqref{eq-MainCond1} and \eqref{eq-MainCond2} are true. Notice that $L_{\kappa,\rho,\rho}(0) = L_{\kappa,\rho}$. The first goal is to show that $L_{\kappa,\rho,\rho'}(\lambda)$ is always invertible and that
it is bounded below by $\frac{g^{2}}{4}\one_{\rho'}$ when $\lambda=0$. The square of $L_{\kappa,\rho,\rho'}(\lambda)$ simplifies to 
$$
L_{\kappa,\rho,\rho'}(\lambda)^{2}
\;=\;
\kappa^{2}\,
\pi_{\rho'}D^{2}\pi_{\rho'}^{*}
\,+\,
\left(\pi_{\rho'}F_{\lambda,\rho}\left(H\otimes\Gamma\right) F_{\lambda,\rho}\pi_{\rho'}^{*}\right)^{2}
\,+\,
\kappa\,\pi_{\rho'}F_{\lambda,\rho}\left[D,\left(H\otimes\sigma_{0}\right)\right]\Gamma F_{\lambda,\rho}\pi_{\rho'}^{*}
\;.
$$
Let us bound the second summand as follows:
\begin{align*}
\big( & \pi_{\rho'} F_{\lambda,\rho}(H\otimes\Gamma) F_{\lambda,\rho}\pi_{\rho'}^{*}\big)^{2}\\
 & =\,\pi_{\rho'}F_{\lambda,\rho}\left(H\otimes\sigma_{0}\right)F_{\lambda,\rho}^{2}\left(H\otimes\sigma_{0}\right)F_{\lambda,\rho}\pi_{\rho'}^{*}\\
 & \geq\,\pi_{\rho'}F_{\lambda,\rho}\left(H\otimes\sigma_{0}\right)F_{\rho}(D)^{2}\left(H\otimes\sigma_{0}\right)F_{\lambda,\rho}\pi_{\rho'}^{*}\\
 & =\,\pi_{\rho'}F_{\lambda,\rho}F_{\rho}(D)\left(H\otimes\sigma_{0}\right)^2F_{\rho}(D)F_{\lambda,\rho}\pi_{\rho'}^{*}+\pi_{\rho'}F_{\lambda,\rho}\left[F_{\rho}(D)\left(H\otimes\sigma_{0}\right),\left[F_{\rho}(D),\left(H\otimes\sigma_{0}\right)\right]\right]F_{\lambda,\rho}\pi_{\rho'}^{*}\\
 & >\,g^{2}\pi_{\rho'}F_{\lambda,\rho}^{2}F_{\rho}(D)^{2}\pi_{\rho'}^{*}+\pi_{\rho'}F_{\lambda,\rho}\left[F_{\rho}(D)\left(H\otimes\sigma_{0}\right),\left[F_{\rho}(D),\left(H\otimes\sigma_{0}\right)\right]\right]F_{\lambda,\rho}\pi_{\rho'}^{*}\\
 & \geq \, g^{2}\pi_{\rho'}F_{\rho}(D)^{4}\pi_{\rho'}^{*}+\pi_{\rho'}F_{\lambda,\rho}\left[F_{\rho}(D)\left(H\otimes\sigma_{0}\right),\left[F_{\rho}(D),\left(H\otimes\sigma_{0}\right)\right]\right]F_{\lambda,\rho}\pi_{\rho'}^{*}
 \;.
\end{align*}
For the special case of $\lambda=0$ one has the better estimate
\begin{align*}
 & \left(\pi_{\rho'}F_{0,\rho}\left(H\otimes\sigma_{0}\right)\Gamma F_{0,\rho}\pi_{\rho'}^{*}\right)^{2}\\
 & \qquad\geq\;g^{2}\pi_{\rho'}F_{\rho}(D)^{2}\pi_{\rho'}^{*}+\pi_{\rho'}F_{\lambda,\rho}\left[F_{\rho}(D)\left(H\otimes\sigma_{0}\right),\left[F_{\rho}(D),\left(H\otimes\sigma_{0}\right)\right]\right]F_{\lambda,\rho}\pi_{\rho'}^{*}
\end{align*}
Furthermore, by spectral calculus of $D$ one has the bound 
$$
\kappa^2\, (D_{\rho'})^2
\;\geq\;
g^2\,\pi_{\rho'}(\one-\Tape(D)^2)\pi_{\rho'}^*
\;,
$$
because the bound  holds for spectral parameters in $[\frac{1}{2}\rho,\rho']$ due to \eqref{eq-MainCond2} and $\one-\Tape(D)^2\leq \one$, while it holds trivially on $[0,\frac{1}{2}\rho]$. Since
$$
\one-F_{\rho}(D)^{2}+F_{\rho}(D)^{4}
\;\geq\;\tfrac{3}{4}\,\one
\;,
$$
it thus follows
$$
L_{\kappa,\rho,\rho'}(\lambda)^{2}
\,>\,
\tfrac{3}{4}\,g^{2}\one_{\rho'}
+
\pi_{\rho'}F_{\lambda,\rho}\left(\left[F_{\rho}(D)\left(H\otimes\sigma_{0}\right),\left[F_{\rho}(D),\left(H\otimes\sigma_{0}\right)\right]\right]+\kappa\left[D,\left(H\otimes\sigma_{0}\right)\right]\Gamma\right)F_{\lambda,\rho}\pi_{\rho'}^{*}
\,,
$$
and in the special case $\lambda=0$,
$$
L_{\kappa,\rho,\rho'}(0)^{2}
\,>\,
g^{2}\,\one_{\rho'}
+
\pi_{\rho'}F_{\lambda,\rho}\left(\left[F_{\rho}(D)\left(H\otimes\sigma_{0}\right),\left[F_{\rho}(D),\left(H\otimes\sigma_{0}\right)\right]\right]+\kappa\left[D,\left(H\otimes\sigma_{0}\right)\right]\Gamma\right)F_{\lambda,\rho}\pi_{\rho'}^{*}
\,.
$$
Finally the error term is bounded using the tapering estimate \eqref{eq-TapeCommu}:
\begin{align*}
\big\|
[\Tape(D)\, H\otimes\sigma_0 & ,[\Tape(D) ,H\otimes\sigma_0 ]]
\,+ \,\kappa[H\otimes\sigma_0 ,D]\Gamma
\big\|
\\
& \;\leq\;
\Big(
2\,\|\Tape(D)\, H\otimes\sigma_0 \|\,8(\rho)^{-1}\,+\,\kappa
\Big)
\|[H\otimes\sigma_0 ,D]\|
\\
&
\;\leq\;
\Big(
\|H\|\,8\,(g)^{-1}\,+\,1\Big)\,\kappa\,\|[H\otimes\sigma_0 ,D]\|
\\
&
\;\leq\;
\|H\|\,9 \,g^{-1}\,\kappa\,\|[H\otimes\sigma_0 ,D]\|
\\
&
\;\leq\;\tfrac{3}{4}\,g^2
\;,
\end{align*}
where the second inequality used   \eqref{eq-MainCond2} as well as $\|\Tape(D) \|=1$, the third one $\|H\|\geq 1$, and finally the last inequality follows from hypothesis \eqref{eq-MainCond1}. Together one infers $L_{\kappa,\rho,\rho'}(\lambda)^{2}>0$ and $L_{\kappa,\rho,\rho'}(0)^{2}\geq\frac{1}{4}\,g^2$. 

\vspace{.1cm}

Finally, let us show that
$$
\Sig\left(L_{\kappa,\rho}\right)
\;=\;
\Sig\left(L_{\kappa',\rho'}\right)
\;,
$$
for pairs $\kappa,\rho$ and $\kappa',\rho'$ in the permitted range of parameters.  Without loss of generality let $\rho\leq\rho'$. Clearly $L_{\kappa,\rho}$ is continuous in $\kappa$, and since any $\kappa$ that is valid
for $\rho$ is valid for $\rho'$ a homotopy argument allows to
reduce to the case $\kappa=\kappa'$, namely one needs to show 
$$
\Sig\left(L_{\kappa,\rho,\rho}(0)\right)
\;=\;
\Sig\left(L_{\kappa,\rho',\rho'}(0)\right)
\;,
$$
when $\rho\leq\rho'$ and \eqref{eq-MainCond1} and \eqref{eq-MainCond2} are true for $\kappa$ and $\rho$. Clearly $L_{\kappa,\rho,\rho}(\lambda)$ is continuous in $\lambda$, so it suffices to prove 
$$
\Sig\left(L_{\kappa,\rho,\rho}(1)\right)
\;=\;
\Sig\left(L_{\kappa,\rho',\rho'}(1)\right)
\;.
$$
Consider 
$$
L_{\kappa,\rho,\rho'}(1)
\;=\;
\kappa\pi_{\rho'}D\pi_{\rho'}^{*}
+\pi_{\rho'}F_{\rho}(D)(H\otimes\Gamma) F_{\rho}(D)\pi_{\rho'}^{*}
\;.
$$
Now $D$ commutes with $\pi_{\rho'}\pi_{\rho'}^*$ so that $L_{\kappa,\rho,\rho'}(1)$ decomposes into a direct sum. Let $\pi_{\rho',\rho}=\pi_{\rho'}\ominus\pi_{\rho}$ be the surjective partial isometry onto $(\Hh\oplus\Hh)_{\rho'}\ominus (\Hh\oplus\Hh)_{\rho}$. Then
$$
L_{\kappa,\rho,\rho'}(1)
\;=\;
L_{\kappa,\rho,\rho}(1)\oplus\,\pi_{\rho',\rho}\,\kappa\,D\,\pi_{\rho',\rho}^{*}
\;.
$$
The signature of $\pi_{\rho',\rho}\,D\,\pi_{\rho',\rho}^{*}$ vanishes so that
$$
\Sig(L_{\kappa,\rho,\rho'}(1))\;=\;\Sig(L_{\kappa,\rho,\rho}(1))
\;.
$$
As $L_{\kappa,\rho,\rho'}(1)$ is continuous in $\rho$, one has $\Sig(L_{\kappa,\rho,\rho'}(1))=\Sig(L_{\kappa,\rho',\rho'}(1))$ by homotopy. 
\hfill $\Box$

\section{Deforming spectral localizer to a fuzzy sphere}
\label{sec-SLocFuzzy}

The first step consists in deforming the invertible selfadjoint operator $H$ into the selfadjoint unitary $\one-2P$ where $P=\chi(H< 0)$. This will be done by the homotopy of invertibles
$$
\lambda\in[0,1]\;\mapsto\;
H(\lambda)
\;=\;
(1-\lambda )H\;+\;\lambda\,(\one-2P)
\;.
$$
One obtains an associated path  $\lambda\in[0,1]\mapsto L_{\kappa,\rho}(\lambda)$ of spectral localizers and has to assure that this path lies within the invertibles, provided $\kappa$ is sufficiently small and $\rho$ is sufficiently large. This follows from Theorem~\ref{theo-invertible} applied to $H(\lambda)$ because $P=p(H)$ for a smooth function $p$ satisfying $\|\widehat{p'}\|_{L^1(\RM)}\leq {2\pi}g^{-1}$ (which can again be constructed explicitly as in Lemma~4 in \cite{LS}). Then by \eqref{eq-TapeCommuGen2}
$$
\| [D,P\otimes\sigma_0]\|
\;=\;
\| [D,p(H\otimes\sigma_0)]\|
\;\leq\;
g^{-1}\,\| [D,H\otimes\sigma_0]\|
\;,
$$
and thus, for $g\leq 1$,
$$
\| [D,H(\lambda)\otimes\sigma_0]\|
\;\leq\;
(1-\lambda+2\lambda g^{-1})\| [D,H\otimes\sigma_0]\|
\;\leq\;
2g^{-1}\| [D,H\otimes\sigma_0]\|
\;.
$$
Replacing $\kappa$ and $\rho$ by $\kappa \frac{g}{2}$ and $\rho\frac{2}{g}$ respectively thus assures that Theorem~\ref{theo-invertible} applies for all $\lambda\in[0,1]$. From now on, we may thus assume that $H$ is a selfadjoint unitary and that $g=1$. 

\vspace{.2cm}

The construction of the fuzzy sphere $(X_1,X_2,X_3)$ appearing in Proposition~\ref{prop-SLocToSphere} will invoke the even and smooth  tapering function $\Tape:\RM\to[0,1]$ already introduced in Section~\ref{sec-Tools} and used in Section~\ref{sec-InvSLoc}. A dual tapering function $\TapeD:\RM\to[0,1]$ is defined by the equation
\begin{equation}
\label{eq-TapeDuality}
\TapeD(x)^4\;+\;\Tape(x)^4\;=\;1
\;.
\end{equation}
The function $\TapeD$ is also even, vanishes on $[-\frac{\rho}{2},\frac{\rho}{2}]$ and is equal to $1$ on $\RM\setminus[-\rho,\rho]$. As again $\TapeD(x)=\TapeDOne(\frac{x}{\rho})$, it satisfies the same bound \eqref{eq-TapeCommu}:
\begin{equation}
\label{eq-TapeDCommu}
\|[\TapeD(D),H\otimes\sigma_0 ]\|
\;\leq\;{C}\,{\rho}^{-1}\,\|[D,H\otimes\sigma_0 ]\|
\;.
\end{equation}
Of course, here $C$ is a different constant. In fact, in the following $C$ will denote different constants (independent of $\rho$, however). We will also need the bound \eqref{eq-TapeCommuGen} for the odd function $G_\rho(x)=\TapeD(x)^2(x^2)^{-\frac{1}{2}}x=G_1(\frac{x}{\rho})$:
\begin{equation}
\label{eq-TapeDCommu2}
\|[\TapeD(D)^2|D|^{-1}D,H\otimes\sigma_0 ]\|
\;\leq\;{C}\,{\rho}^{-1}\,\|[D,H\otimes\sigma_0 ]\|
\;.
\end{equation}
From now on, we will heavily use that $D_0$ is normal. Then $\Hh_\rho=\Hh_{\rho,\pm}$ and $\pi_{\rho,+}=\pi_{\rho,-}$. We will use the notation $T_\rho=\pi_{\rho,\pm}T \pi_{\rho,\pm}$ also for restrictions of operators $T$ on $\Hh$. Also, let us set $D_1=\frac{1}{2}(D_0+D_0^*)$ and $D_2=\frac{1}{2\imath}(D_0-D_0^*)$. Then $[D_1,D_2]=0$. Furthermore, let us introduce a non-negative operator $R$ on $\Hh$ by 
\begin{equation}
\label{eq-RDef}
R^2
\;=\;
D_1^2+D_2^2
\;.
\end{equation}
One has $D^2=R^2\otimes\sigma_0$ and $|D|=R\otimes\sigma_0$. For the even function $\TapeD$ also $\TapeD(D)=\TapeD(R)\otimes\sigma_0$. Thus \eqref{eq-TapeDCommu2} implies
$$
\|[\TapeD(R)^2R^{-1}(D_1\pm\imath\,D_2),H]\|
\;\leq\;{C}\,{\rho}^{-1}\,\|[D,H\otimes\sigma_0 ]\|
\;,
$$
and therefore also, for $i=1,2$,
\begin{equation}
\label{eq-TapeDCom}
\|
[\TapeD(R)^2R^{-1}D_i,H]\|
\;\leq\;2\,{C}\,{\rho}^{-1}\,\|[D,H\otimes\sigma_0 ]\|
\;.
\end{equation}

Let us now introduce $3$ selfadjoint operators on the finite dimensional Hilbert space $\Hh_\rho$:
\begin{align}
& 
X_1\;=\;\TapeD(R)\,R^{-\frac{1}{2}}\,D_{1,\rho}\,R^{-\frac{1}{2}}\,\TapeD(R)\;,
\nonumber
\\
&
X_2\;=\;\TapeD(R)\,R^{-\frac{1}{2}}\,D_{2,\rho}\,R^{-\frac{1}{2}}\,\TapeD(R)\;,
\label{eq-FuzSpheConcrete}
\\
& X_3\;=\;\Tape(R)\,H_\rho\,\Tape(R)\;.
\nonumber
\end{align}
Each of the $X_j$, $j=1,2,3$, depends on $\rho$ and will also be seen as a net $X_j=(X_{j,\rho})$ in the following.

\begin{lemma}
Let $D_0$ be normal. Then $(X_1,X_2,X_3)$ is a fuzzy sphere of width of order $\rho^{-1}$.
\end{lemma}

\noindent {\bf Proof.} First note that $X_j=\TapeD(R)^2 R^{-1}\,D_{2,\rho}$. Due to $H^2=\one$,
\begin{align*}
X_1^2\,+\,X_2^2\,+\,X_3^2
\;=\;
&
\TapeD(R)^4\,R^{-2}\,(D_{1,\rho}^2\,+\,D_{2,\rho}^2)
\;+\;
\Tape(R)^4\,+\,\Tape(R)\one_\rho[H,\Tape(R)^2]H\Tape(R)
\;.
\end{align*}
Now, using $\|\Tape(R)\|\leq 1$ and \eqref{eq-TapeCommu},
$$
\|[H,\Tape(R)^2]\|\;\leq\;2\,\|[H,\Tape(R)]\|\;\leq\;{16}\,{\rho}^{-1}\,\|[H\otimes\sigma_0,D]\|
\;,
$$
and the last factor can be bounded by \eqref{eq-MainCond1}, with a bound that is independent of $\rho$. Using \eqref{eq-TapeDuality}
$$
\|
X_1^2\,+\,X_2^2\,+\,X_3^2\,-\,\one\|
\;\leq\;16\,\|H\|\;\|[H\otimes\sigma_0,D]\|\,\rho^{-1}
\;\leq\;C\,\rho^{-1}
\;,
$$
for some constant $C$. Furthermore, the commutator $[X_1,X_2]$ vanishes, and the two others $[X_1,X_3]$ and $[X_2,X_3]$ can be bounded by a constant times $\rho^{-1}$ by using \eqref{eq-TapeDCom}.
\hfill $\Box$

\vspace{.2cm}

\noindent {\bf Proof} of Proposition~\ref{prop-SLocToSphere}. The basic idea of the argument is the same as in the previous Section~\ref{sec-InvSLoc}. Let us set
$$
\TapeD(R,\lambda)\;=\;(1-\lambda)\,\kappa^{\frac{1}{2}}\,{R^{\frac{1}{2}}}\,\one_\rho\;+\;\lambda\,\TapeD(R)\;,
\qquad
\Tape(R,\lambda)\;=\;(1-\lambda)\,\one_\rho\;+\;\lambda\,\Tape(R)\;.\,
$$
and then
\begin{align*}
&
X_1(\lambda)\;=\;\TapeD(R,\lambda)\,{R^{-1}}\,D_{1,\rho}\,\TapeD(R,\lambda)\;,
\\
&
X_2(\lambda)\;=\;\TapeD(R,\lambda)\,{R^{-1}}\,D_{2,\rho}\,\TapeD(R,\lambda)\;,
\\
& 
X_3(\lambda)\;=\;\Tape(R,\lambda)\,H_\rho\,\Tape(R,\lambda)\;,
\end{align*}
and finally from these operators
$$
L_{\kappa,\rho}(\lambda)
\;=\;
\sum_{j=1,2,3} X_j(\lambda)\otimes\sigma_j
\;.
$$
Then $L_{\kappa,\rho}(0)=L_{\kappa,\rho}$ is the spectral localizer and $L_{\kappa,\rho}(1)$ is the selfadjoint associated to the fuzzy sphere \eqref{eq-FuzSpheConcrete}. The proof is hence concluded by showing that $L_{\kappa,\rho}(\lambda)$ is invertible for all $\lambda\in[0,1]$ and for $\rho$ sufficiently large.

\vspace{.1cm}

Again one starts by calculating $L_{\kappa,\rho}(\lambda)^2$ as in \eqref{eq-LSquare}. Due to the $[D_1,D_2]=0$, the summand $X_1(\lambda)^2+X_2(\lambda)^2$ can readily be calculated:  
$$
X_1(\lambda)^2+X_2(\lambda)^2
\;=\;
(D_{1,\rho}^2\,+\,D_{2,\rho}^2)\,{R^{-2}}\,\TapeD(R,\lambda)^4
\;=\;
\TapeD(R,\lambda)^4
\;.
$$
Furthermore, one summand can be bounded similar as in the proof of  Theorem~\ref{theo-invertible}:
\begin{align*}
X_3(\lambda)^2
& \;=\; \Tape(R,\lambda) H\Tape(R,\lambda)^2 H\Tape(R,\lambda) 
\\
& 
\;\geq\;
\Tape(R,\lambda) H\Tape(R)^2 H\Tape(R,\lambda)
\\
& 
\;=\;
\Tape(R,\lambda)\Tape(R)^2\Tape(R,\lambda)
\,+\,
\Tape(R,\lambda)
\big[[H,\Tape(R) ],\Tape(R) \,H\big]
\Tape(R,\lambda)
\\
& 
\;\geq\;
\Tape(R)^4
\,+\,
\Tape(R,\lambda)
\big[[H,\Tape(R) ],\Tape(R) \,H\big]
\Tape(R,\lambda)
\;,
\end{align*}
where $H^2=\one$ was used. As $\|\Tape(R,\lambda)\|\leq 1$ and $\|[H,\Tape(R) ]\|\leq C\rho^{-1}$ by \eqref{eq-TapeCommu},
$$
\|
X_3(\lambda)^2
\,-\,
\Tape(R)^4
\|
\;\leq\;
C\,\rho^{-1}
\;,
$$
uniformly in $\lambda\in[0,1]$.  Another summand vanishes as $[X_1(\lambda),X_2(\lambda)]=0$ and, for $i=1,2$,
\begin{align*}
\|[X_i(\lambda),X_3(\lambda)]\|
& \;=\;
\|
\Tape(R,\lambda) [\TapeD(R,\lambda)^2\,{R^{-1}}\,D_{i,\rho},H_\rho]
\Tape(R,\lambda)
\|
\\
&
\;\leq\;
\|
 [\TapeD(R,\lambda)^2\,{R^{-1}}\,D_{i,\rho},H_\rho]
\|
\\
&
\;\leq\;
(1-\lambda)^2\kappa
\|
[D_{i,\rho},H_\rho]
\|
\;+\;
2(1-\lambda)\lambda\kappa^{\frac{1}{2}}
\|
 [\TapeD(R,\lambda)\,{R^{-\frac{1}{2}}}\,D_{i,\rho},H_\rho]
\|
\\
& \;\;\;\;\;\;\;
\;+\;
\lambda^2 
\|[\TapeD(R)^2\,{R^{-1}}\,D_{i,\rho},H_\rho]\|
\\
&
\;\leq\;
(1-\lambda)^2\kappa
\|
[D_{i},H]
\|
\;+\;
2(1-\lambda)\lambda\kappa^{\frac{1}{2}}
\|\TapeD(R)\,{R^{-\frac{1}{2}}}\|\,
\|
 [D_{i,\rho},H_\rho]
\|
\\
& \;\;\;\;\;\;\;
\;+\;
2(1-\lambda)\lambda\kappa^{\frac{1}{2}}
\|[\TapeD(R)\,{R^{-\frac{1}{2}}},H_\rho]\|\,
\|
D_{i,\rho}
\|
\;+\;
\lambda^2 
\|[\TapeD(R)^2\,{R^{-1}}\,D_{i},H]\|
\;.
\end{align*}
Now $\TapeD$ vanishes on $[-\frac{\rho}{2},\frac{\rho}{2}]$ so that $\|\TapeD(R)\,{R^{-\frac{1}{2}}}\|\leq 2^{\frac{1}{2}}\rho^{-\frac{1}{2}}$. Also, $\TapeD(R)\,{R^{-\frac{1}{2}}}=\rho^{-\frac{1}{2}}G_\rho(R)$ for some function $G_\rho$ for which \eqref{eq-TapeCommuGen} holds. Thus, with $\|D_{i,\rho}\|\leq\rho$,
\begin{align*}
\|[X_i(\lambda),X_3(\lambda)]\|
& \;\leq\;
(1-\lambda)^2\kappa\,(18\kappa)^{-1}\,+\,2(1-\lambda)\lambda\kappa^{\frac{1}{2}}
\Big(2^{\frac{1}{2}}\rho^{-\frac{1}{2}}\,(18\kappa)^{-1}
+\rho^{-\frac{1}{2}}\,C
\Big)
\,+\,
\lambda^2 C\,(\rho)^{-1}
\\
&
\;\leq\;
(1-\lambda)^2\tfrac{1}{18}\,+\,C'\,\rho^{-\frac{1}{2}}
\;,
\end{align*}
for some constant $C'$ depending on $\kappa$, but not on $\lambda$.  Then replacing all the above shows
$$
L_{\kappa,\rho}(\lambda)^2
\;\geq\;
\Tape(R)^4\;+\;\TapeD(R_\rho,\lambda)^4
\;-\;
\Big((1-\lambda)^2\tfrac{1}{18}\,+\,C''\,\rho^{-\frac{1}{2}}\Big)
\;.
$$
It hence only remains to show that the r.h.s.\ remains positive. By functional calculus of $R_\rho$, this is merely a statement about the functions involved. For spectral parameters $r\in[0,\frac{\rho}{2}]$, the claim is obvious because then $\Tape(r)=1$. For $r\in[\frac{\rho}{2},\rho]$,
\begin{align*}
\Tape(r)^4\,+\,\TapeD(r,\lambda)^4\,-\,(1-\lambda)^2\tfrac{1}{18}
& \;\geq\;
\lambda^4\Tape(r)^4\,+\,\big(\kappa^2r^2(1-\lambda)^4+\lambda^4\TapeD(r)^4\big)\,-\,(1-\lambda)^2\tfrac{1}{18}
\\
&
\;\geq\;
\lambda^4\,+\,\frac{\rho^2}{4}\,\kappa^2(1-\lambda)^4\,-\,(1-\lambda)^2\tfrac{1}{18}
\\
&
\;\geq\;
\lambda^4\,+\,(1-\lambda)^4\,-\,(1-\lambda)^2\tfrac{1}{18}
\;,
\end{align*}
the latter due to \eqref{eq-MainCond2}. This is strictly larger than $C''\,\rho^{-\frac{1}{2}}$ for $\rho$ sufficiently large.
\hfill $\Box$

\section{Image of the index map as a fuzzy sphere}
\label{sec-IndMap}

In this section, it will be shown how to modify the Fredholm operator $T=P\,F\,P\,+\,(\one-P)$ appearing in the even index pairing of Theorem~\ref{theo-pairings} within the set of Fredholm operators to an operator $A$ satisfying hypothesis of Theorem~\ref{theo-IndSphere}, with $\Kk$ and $\Bb$ being the compact and bounded operators on $\Hh$. We will already assume $[D_1,D_2]=0$. Then $R^2=|D_1+\imath\,D_2|^2=D_1^2+D_2^2$ as in \eqref{eq-RDef}, and $[D_i,R]=0$ for $i=1,2$. First, let us decompose $T=T_1+\imath\,T_2$ into real and imaginary part:
$$
T_1\;=\;
P\,R^{-1}\,D_1 \,P\;+\;(\one-P)
\;,
\qquad
T_2\;=\;
P\,R^{-1}\,D_2\,P
\;.
$$
Now $\TapeD(R)-\one$ is compact. Hence setting 
\begin{equation}
A_1\,=\,
P\,\TapeD(R)^2\,R^{-1}\,D_1\,P\;+\;(\one-P)\TapeD(R)^2(\one-P)
\;,
\qquad
A_2\,=\,
P\,\TapeD(R)^2\,R^{-1}\,D_2\,P
\;,
\label{eq-A1A2}
\end{equation}
as well as $A=A_1+\imath\,A_2$, one clearly has $\pi(A)=\pi(T)$ where $\pi$ is the projection onto the Calkin algebra. Again, $A_1$, $A_2$ and $A$ depend on $\rho$ and will be seen as nets with index $\rho$. As $H=\one-2P$, \eqref{eq-TapeDCom} implies
$$
[A_1,A_2]
\;\sim_\rho\;
0
\;.
$$
%
%
Thus choosing $\rho$ sufficiently large, one can assure \eqref{eq-LiftCommEst} so that Theorem~\ref{theo-IndSphere} can be applied. This provides a fuzzy sphere $(Y_1,Y_2,Y_3)$ in $\Kk$ which provides the image of the index map. Again this fuzzy sphere depends on $\rho$. The following result spells out a slightly modified version $(Y'_1,Y'_2,Y'_3)$ of this fuzzy sphere.

\begin{proposi}
\label{prop-IndSphere}
Let $\IndOne:[0,1]\to [0,1]$ and $\IndTwo:[0,1]\to [-1,1]$ be the two smooth functions appearing in {\rm Theorem~\ref{theo-IndSphere}}, notably satisfying \eqref{eq-IndFunct}. Set
\begin{align*}
& Y'_1
\;=\;
P\,\IndOne(\TapeD(R)^2)^2\,\TapeD(R)^2\,R^{-1}\,D_1\,P
\,+\,
(\one-P)\IndOne(\TapeD(R)^2)^2\TapeD(R)^2(\one-P)
\;,
\\
&
Y'_2
\;=\;
-\,P\,\IndOne(\TapeD(R)^2)^2\,\TapeD(R)^2\,R^{-1}\,D_2\,P
\;,
\\
& 
Y'_3\;=\;\IndTwo(\TapeD(R)^2)
\;.
\end{align*}
Then  $(Y'_1,Y'_2,Y'_3)$ form a fuzzy sphere in $\Kk$ of width of order of $\rho^{-2}$ and the image of the index paring $T$ as given in \eqref{eq-EvenPair} under the $K$-theoretic index is 
$$
\Ind [\pi(T)]_1
\;=\;
\Big[
\sum_{j=1,2,3} Y'_j\otimes\sigma_j
\Big]_0
\;.
$$
\end{proposi}

\noindent {\bf Proof.} The proof consists in proving that $(Y_1,Y_2,Y_3)$ given by \eqref{eq-YSphere} with $A_1$ and $A_2$ as in \eqref{eq-A1A2} is equal to $(Y'_1,Y'_2,Y'_3)$, up to errors of order $\rho^{-1}$. Let us begin by calculating $B^2=A_1^2+A_2^2$:
\begin{align*}
B^2
\;=\;
& P\,\TapeD(R)^2\,R^{-1}\,D_1 \,P\,\TapeD(R)^2\,R^{-1}\,D_1 \,P\,+\,
(\one-P)\TapeD(R)^2(\one-P)\TapeD(R)^2(\one-P)
\\
& \,+\,
P\,\TapeD(R)^2\,R^{-1}\,D_2\, P\TapeD(R)^2\,R^{-1}\,D_2\,P
\;.
\end{align*}
Note that
\begin{align*}
B^2
\;\leq\;
& P\,\TapeD(R)^2\,R^{-1}\,D_1 \,\TapeD(R)^2\,R^{-1}\,D_1 \,P\,+\,
(\one-P)\TapeD(R)^2\TapeD(R)^2(\one-P)
\\
& \,+\,
P\,\TapeD(R)^2\,R^{-1}\,D_2\,\TapeD(R)^2\,R^{-1}\,D_2\,P
\\
\;=\;
&
P\,\TapeD(R)^4\,P\;+\;(\one-P)\TapeD(R)^4(\one-P)
\;,
\end{align*}
and, in particular, $B^2\leq \one$. More precisely, one has due to \eqref{eq-TapeDCom}
$$
B^2
\,\sim_\rho \,
\big(P\,\TapeD(R)^4\,P\,+\,(\one-P)\TapeD(R)^4(\one-P)\big)
\;.
$$
Again by \eqref{eq-TapeDCommu} one also has $P\,\TapeD(R)^4(\one-P)\sim_\rho 0$ so that $B^2\sim_\rho 
\TapeD(R)^4$. Applying Lemma~\ref{lem-RhoBehave}(iii) one therefore finally obtains for the roots
$$
B
\,\sim_{\rho^2} \,
\TapeD(R)^2
\;.
$$
As $\IndTwo$ is smooth, this also implies by Lemma~\ref{lem-RhoBehave}(i) that $Y_3-Y'_3=\IndTwo(B)-\IndTwo(\TapeD(R)^2)$ is of order $\rho^{-2}$. Furthermore, $\IndOne(B)-\IndOne(\TapeD(R)^2)$ is of order $\rho^{-2}$ so that
$$
Y_2\,\sim_{\rho^2} \,-\,\IndOne(\TapeD(R)^2)\,A_2\,\IndOne(\TapeD(R)^2)
\;.
$$
As now also $[\IndOne(\TapeD(R)^2),P]\sim_{\rho} 0$, one has
$$
\IndOne(\TapeD(R)^2)\,A_2\,\IndOne(\TapeD(R)^2)\;\sim_{\rho^2} \;
P\,\IndOne(\TapeD(R)^2)^2\,\TapeD(R)^2\,R^{-1}\,D_2\,P
\;.
$$
This means that $Y_2\sim_{\rho^2} Y'_2$. Similarly, $Y_1\sim_{\rho^2}  Y'_1$.
\hfill $\Box$

\section{Deforming the fuzzy sphere $(X_1,X_2,X_3)$}
\label{sec-DeformFuzzy}

Let us first prove Proposition~\ref{prop-SphereToSphere} which is the explicit analysis of a degree $1$ map $\Ff$ on the $2$-sphere. This map $(z_1,z_2,z_3)=\Ff(x_1,x_2,x_3)$ is of the form
$$
\begin{pmatrix} z_1 \\ z_2 \\ z_3 \end{pmatrix}
\;=\;
\begin{pmatrix} 
\IndOneD(x_3)^2 x_1\,+\,\IndOneD(-x_3)^2(1-x_3^2)^{\frac{1}{2}}
\\
-\,\IndOneD(x_3)^2x_2
\\
\IndTwoD(x_3)
\end{pmatrix}
\;=\;
\begin{pmatrix} 
\IndOneD(x_3)^2 x_1\,+\,\chi(x_3<0)\big(1-\IndTwoD(x_3)^2\big)^{\frac{1}{2}}
\\
-\,\IndOneD(x_3)^2x_2
\\
\IndTwoD(x_3)
\end{pmatrix}
\;,
$$
where $\IndOneD:[-1,1]\to [0,1]$ and $\IndTwoD:[-1,1]\to [-1,1]$ are continuous functions satisfying the conditions \eqref{eq-IndFunctD}, and $\chi(x_3<0)$ denotes the indicator function on negativ $x_3$. Beforehand, let us give a particular realization:
\begin{equation}
\label{eq-SpecialChoice}
\IndTwoD(x)
\;=\;
1\,-\,2\,|x|
\;,
\qquad
\IndOneD(x)
\;=\;
\left\{
\begin{array}{cc}
0 \;, & x\leq 0 \;, \\
\big(4x(1+x)^{-1}\big)^{\frac{1}{4}}\;, & x\geq 0
\;.
\end{array}
\right.
\end{equation}
Let us also note that the equation $(1-x^2)\IndOneD(x)^4+\IndTwoD(x)^2=1$ fixes $\IndOneD$ if $\IndTwoD$ is given, or inversely it fixes $\IndTwoD$ if $\IndOneD$ is given. For example, the even function $\IndTwoD$ with $\IndTwoD(0)=1=-\IndTwoD(1)$ determines $\IndOneD(x)$ for $x\geq 0$ due to $(1-x^2)\IndOneD(x)^4+\IndTwoD(x)^2=1$, and $\IndOneD(-x)=0$ for $x\geq 0$ is imposed anyway. While the above choice of $\IndTwoD$ is continuous, it is not differentiable. It may, however, be useful to have a concrete example to visualize the constructions below. Differentiable choices with nice behavior at the boundary points $0$ and $1$  are obtained if $\IndTwoD(x)=1-cx^{\alpha}+\mbox{o}(x^{\alpha})$ with $\alpha\geq 4$ and $\IndTwoD(x)=-1+c(x-1)^{\beta}+\mbox{o}((x-1)^{\beta})$ with $\beta\geq 5$. Then analysis of $(1-x^2)\IndOneD(x)^4+\IndTwoD(x)^2=1$ shows that also $\IndOneD$ has then bounded derivatives at $0$ and $1$.

\vspace{.2cm}

\noindent {\bf Proof} of Proposition~\ref{prop-SphereToSphere}:  The map $\Ff$ indeed maps $\SM^2$ to itself because, due to $\IndOneD(-x)\IndOneD(x)=0$ for all $x\in[-1,1]$ and the identities \eqref{eq-IndFunctD},
\begin{align*}
z_1^2+z_2^2+z_3^2
&
\;=\;
\IndOneD(x_3)^4 (x_1^2+x_2^2)\,+\,\IndOneD(-x_3)^4(1-x_3^2)
\,+\,\IndTwoD(x_3)^2
\\
&
\;=\;
(\IndOneD(x_3)^4 \,+\,\IndOneD(-x_3)^4)(1-x_3^2)
\,+\,\IndTwoD(x_3)^2
\\
& \;=\;1
\;.
\end{align*}
It ought to be stressed that the mapping is surjective, but highly non-injective. Actually, a whole half-sphere is mapped to just one arch. Nevertheless, the map is continuous and thus has a mapping degree. It is most easy to calculate this degree at a regular point by use of differential topology ({\it e.g.} \cite{Hir}). For example, let us suppose that $\IndOneD$ and $\IndTwoD$ are given by \eqref{eq-SpecialChoice} and then consider the point $(z_1,z_2,z_3)=(0,1,0)$. From $z_3=\IndTwoD(x_3)=0$ one infers $x_3=\pm \frac{1}{2}$; as $z_2\geq 0$ it follows that actually $x_3=\frac{1}{2}$; as $z_1=0$ one then deduces $x_1=0$; finally $1=-\IndOneD(\frac{1}{2})^2x_2$ implies $x_2=-\frac{\sqrt{3}}{2}$. Hence $(z_1,z_2,z_3)=(0,1,0)$ has only one preimage $(x_1,x_2,x_3)=(0,-\frac{\sqrt{3}}{2}, \frac{1}{2})$. As the mapping degree is equal to the sum of signs of the determinants of Jacobians over all preimages, it can only be $1$ or $-1$. Calculating the derivates at $(x_1,x_2,x_3)=(0,-\frac{\sqrt{3}}{2}, \frac{1}{2})$ shows that the mapping degree is actually $1$. Now for any other function $\IndTwoD$ one can consider the homotopy $\Lambda\in[0,1]\mapsto\IndTwoD_\Lambda(x)=(1-\Lambda)(1-2|x|)+\Lambda \IndTwoD(x)$ and (uniquely) associated functions $\IndOneD_\Lambda$, during which the mapping degree does not change. 

\vspace{.1cm}

Now, as the mapping degree of $\Ff$ is equal to $1$, it is well-known that $\Ff$ is homotopic to the identity. To write out an explicit homotopy $\lambda\in[0,1]\mapsto\Ff_\lambda$ of differentiable maps, let us set $\IndTwoD_\lambda(x)=\IndTwoD\big(1-\frac{1}{2-\lambda}(1-x)\big)$ and then define $\IndOneD_\lambda$ by $\IndOneD_\lambda(x)=0$ for $x\leq \lambda-1$ and by $(1-x^2)\IndOneD_\lambda(x)^4+\IndTwoD_\lambda(x)^2=1$ for $x\geq \lambda-1$. Then set
\begin{equation}
\label{eq-FuzzyHom}
\Ff_\lambda\begin{pmatrix} x_1 \\ x_2 \\ x_3 \end{pmatrix}
\;=\;
\begin{pmatrix} 
\IndOneD_\lambda(x_3)^2 x_1\,+\,\chi(x_3< \lambda-1)\,\big(1-\IndTwoD_\lambda(x_3)^2\big)^{\frac{1}{2}}
\\
-\,\IndOneD_\lambda(x_3)^2x_2
\\
\IndTwoD_\lambda(x_3)
\end{pmatrix}
\;,
\end{equation}
where $\chi(x_3<\lambda-1)$ denotes the characteristic function on $x_3<\lambda-1$. As $\IndTwoD_\lambda(\lambda-1)=1$, the map $\Ff_\lambda$ is continuous, and actually even differentiable at the discontinuity of the characteristic function. Also, by construction $\Ff_\lambda$ sends $\SM^2$ to $\SM^2$.  Moreover, $\IndTwoD_0:[-1,1]\to[-1,1]$ can be smoothly deformed into minus the identity ({\it e.g.} by a linear homotopy) and then $\IndOneD_0$ is accordingly deformed to the function identically equal to $1$. Thus  $\Ff_0$ is homotopic to the map $(x_1,x_2,x_3)\mapsto (x_1,-x_2,-x_3)$, which after a rotation in the 2-3 plane by $\pi$ is seen to be homotopic to the identity.
\hfill $\Box$

\vspace{.2cm}

Now the homotopy $\lambda\in[0,1]\mapsto \Ff_\lambda$ of differentiable maps on $\SM^2$ is used to obtain a homotopy of fuzzy spheres.

\begin{proposi}
\label{prop-FuzSphereHomotopy} For $\rho$ sufficiently large, the fuzzy sphere $(X_1,X_2,X_3)$ given in {\rm Proposition~\ref{prop-SLocToSphere}} can be deformed within the set of fuzzy sphere on $\Kk$ to the fuzzy sphere $(Z_1,Z_2,Z_3)$ defined by \eqref{eq-MappedFuzzy}. In particular, both spheres define the same element of $K_0(\Kk)$ via {\rm Proposition~\ref{eq-SphereK0}}. 
\end{proposi}

\noindent {\bf Proof.} The homotopy $\lambda\in[0,1]\mapsto(Z_{1,\lambda},Z_{2,\lambda},Z_{3,\lambda})$ is defined using the maps \eqref{eq-FuzzyHom}:
\begin{align*}
& Z_{1,\lambda}\,=\,\IndOneD_\lambda(X_3)X_1\IndOneD_\lambda(X_3)\,+\,\chi(X_3<\lambda-1)\big(\one-\IndTwoD_\lambda(X_3)^2\big)^{\frac{1}{2}}\;,
\\
& Z_{2,\lambda}\,=\,-\,\IndOneD_\lambda(X_3)X_2\IndOneD_\lambda(X_3)\;,
\\
& Z_{3,\lambda}\,=\,\IndTwoD_\lambda(X_3)
\;.
\end{align*}
As $\IndOneD_\lambda(X_3)\chi(X_3<\lambda-1)=0=\chi(X_3<\lambda-1)\IndOneD_\lambda(X_3)$ and the commutators $[\IndOneD_\lambda(X_3),X_i]$ and $[\IndTwoD_\lambda(X_3),X_i]$ are of the order $\rho^{-1}$ for $i=1,2$, it follows from the commutative identities that $(Z_{1,\lambda},Z_{2,\lambda},Z_{3,\lambda})$ is indeed a fuzzy sphere of width $\rho^{-1}$ for all $\lambda\in [0,1]$. 
\hfill $\Box$

\section{Comparing fuzzy spheres}
\label{sec-CompFuzzy}

In this section, we complete the proof of Theorem~\ref{theo-MainRes} by showing the fuzzy sphere $(Y'_1,Y'_2,Y'_3)$ given in Proposition~\ref{prop-IndSphere} is homotopic to $(Z_1,Z_2,Z_3)$ given in \eqref{eq-MappedFuzzy} with $(X_1,X_2,X_3)$ as in \eqref{eq-FuzSpheConcrete} provided that the functions $\IndOne$, $\IndTwo$  and $\IndOneD$, $\IndTwoD$ are related via \eqref{eq-duality} and $\rho$ is sufficiently large.

\vspace{.2cm}

\noindent {\bf Proof} of Theorem~\ref{theo-MainResInd}: Let us begin by expressing $(Z_1,Z_2,Z_3)$ in terms of $R$, $D_1$ and $D_2$ by replacing \eqref{eq-FuzSpheConcrete} in  \eqref{eq-MappedFuzzy}. This requires the evaluation of $\IndOneD(\Tape(R)H_\rho\Tape(R))$ and $\IndTwoD(\Tape(R)H_\rho\Tape(R))$. Here $\IndOneD$ and $\IndTwoD$ are smooth real functions on $[-1,1]$ specified in Proposition~\ref{prop-SphereToSphere}, and $\Tape(R)H_\rho\Tape(R)=\Tape(R)H\Tape(R)$ is a selfadjoint operator (of finite dimensional  range) of norm less than or equal to $1$. Using $H=P-(\one-P)$ and the commutator estimate \eqref{eq-TapeCommu} one has
\begin{align*}
\Tape(R)H_\rho\Tape(R)
&
\,\;=\;\;
\Tape(R)\,P\,\Tape(R)\;-\;\Tape(R)\,(\one-P)\,\Tape(R)
\\
&
\;\sim_\rho\;
P\,\Tape(R)^2\,P\;+\;(\one-P)\,(-\Tape(R)^2)\,(\one-P)
\;.
\end{align*}
Let us stress that the operator on the l.h.s.\ is strictly local (supported by $\Hh_\rho$) while the one on the r.h.s.\ is not. On the other, the two summands on the r.h.s.\ are orthogonal, which is not true on the l.h.s. Squaring leads to
\begin{align*}
\big(\Tape(R)H_\rho\Tape(R)\big)^2
&
\,\;=\;\;
P\,\Tape(R)^2\,P\,\Tape(R)^2\,P\;+\;(\one-P)\,\Tape(R)^2\,(\one-P)\,\Tape(R)^2\,(\one-P)
\\
&
\;\sim_\rho\;
P\,\Tape(R)^4\,P\;+\;(\one-P)\,\Tape(R)^4\,(\one-P)
\\
&
\;\sim_\rho\;
\Tape(R)^4
\;.
\end{align*}
As $\IndTwoD$ is smooth, one now gets with Lemma~\ref{lem-RhoBehave}(ii)
\begin{align*}
\IndTwoD(\Tape(R)H_\rho\Tape(R))
&
\;\sim_\rho\;
\IndTwoD\big(P\,\Tape(R)^2\,P\;+\;(\one-P)\,(-\Tape(R)^2)\,(\one-P)\big)
\\
&
\;=\;\;\,
\IndTwoD\big(P\,\Tape(R)^2\,P\big)\;+\;\IndTwoD\big((\one-P)\,(-\Tape(R)^2)\,(\one-P)\big)
\\
&
\;\sim_\rho\;
P\,\IndTwoD(\Tape(R)^2)\,P\;+\;(\one-P)\,\IndTwoD(-\Tape(R)^2)\,(\one-P)
\\
&
\;=\;\;
P\,\IndTwoD(\Tape(R)^2)\,P\;+\;(\one-P)\,\IndTwoD(\Tape(R)^2)\,(\one-P)
\\
&
\;\sim_\rho\;
\IndTwoD(\Tape(R)^2)
\;.
\end{align*}
Similarly
\begin{align*}
\IndOneD(\Tape(R)H_\rho\Tape(R))
&
\;\sim_\rho\;
P\,\IndOneD(\Tape(R)^2)\,P\;+\;(\one-P)\,\IndOneD(-\Tape(R)^2)\,(\one-P)
\\
&
\;=\;
\;P\,\IndOneD(\Tape(R)^2)\,P
\;,
\end{align*}
and
$$
\IndOneD(-\Tape(R)H_\rho\Tape(R))
\;\sim_\rho\;
(\one-P)\,\IndOneD(\Tape(R)^2)\,(\one-P)
\;.
$$
Replacing $X_3=\Tape(R)H_\rho\Tape(R)$ and the above into \eqref{eq-MappedFuzzy} leads to
\begin{align}
& Z_1\,\sim_\rho\,
P\,\IndOneD(\Tape(R)^2)\,P\,X_1\,P\,\IndOneD(\Tape(R)^2)\,P 
\nonumber
\\
& \;\;\;\;\;\;\;\;\;\;\;\;\;\; \,+\,
(\one-P)\,\IndOneD(\Tape(R)^2)\,(\one-P)\,(\one-\Tape(R)^4)^{\frac{1}{2}}\,(\one-P)\,\IndOneD(\Tape(R)^2)\,(\one-P)\;,
\nonumber
\\
& Z_2\,\sim_\rho\,-\,P\,\IndOneD(\Tape(R)^2)\,P\,X_2\,P\,\IndOneD(\Tape(R)^2)\,P\;,
\nonumber
\\
& Z_3\,\sim_\rho\,\IndTwoD(\Tape(R)^2)
\;.
\nonumber
\end{align}
Now the $P$ and $\one-P$ can be commuted to the outside, up to errors of the order of $\rho^{-1}$. As $X_i=\TapeD(R){R^{-1}}D_{i,\rho}\TapeD(R)$, one thus gets
\begin{align}
& Z_1\,\sim_\rho\,
P\,\IndOneD(\Tape(R)^2)\,\TapeD(R){R^{-1}}D_{1,\rho}\TapeD(R)\,\IndOneD(\Tape(R)^2)\,P 
\nonumber
\\
& \;\;\;\;\;\;\;\;\;\;\;\;\;\; \,+\,
(\one-P)\,\IndOneD(\Tape(R)^2)\,(\one-\Tape(R)^4)^{\frac{1}{2}}\,\IndOneD(\Tape(R)^2)\,(\one-P)\;,
\nonumber
\\
& Z_2\,\sim_\rho\,-\,P\,\IndOneD(\Tape(R)^2)\,\TapeD(R){R^{-1}}D_{2,\rho}\TapeD(R)\,\IndOneD(\Tape(R)^2)\,P\;,
\nonumber
\\
& Z_3\,\sim_\rho\,\IndTwoD(\Tape(R)^2)
\;.
\nonumber
\end{align}
Using $[D_i,R]=0$ and \eqref{eq-TapeDuality}, one hence has
\begin{align}
& Z_1\,\sim_\rho\,
P\,\IndOneD(\Tape(R)^2)^2\,\TapeD(R)^2\,{R^{-1}}D_1\,P 
\,+\,
(\one-P)\,\IndOneD(\Tape(R)^2)^2\,
{\TapeD(R)^2\,}
(\one-P)\;,
\nonumber
\\
& Z_2\,\sim_\rho\,-\,P\,\IndOneD(\Tape(R)^2)^2\,\TapeD(R)^2\,{R^{-1}}\,D_2\,P\;,
\label{eq-MappedFuzzy3}
\\
& Z_3\,\sim_\rho\,\IndTwoD(\Tape(R)^2)
\;.
\nonumber
\end{align}
Using the duality equation \eqref{eq-duality} combined with \eqref{eq-TapeDuality}, one finds $\IndOneD(\Tape(R)^2)=\IndOne(\TapeD(R)^2)$ and $\IndTwoD(\Tape(R)^2)=\IndTwo(\TapeD(R)^2)$. Therefore comparing with Proposition~\ref{prop-IndSphere} shows that \eqref{eq-MappedFuzzy3} merely says $Z_i\sim_{\rho} Y'_i$ for $i=1,2,3$. Combined with Proposition~\ref{prop-IndSphere} this concludes the proof.
\hfill $\Box$

\vspace{.4cm}

\noindent{\bf Acknowledgments:} The authors thank the Simons Foundatin (CGM	419432), the NSF (DMS 1700102) and the DFG (SCHU 1358/3-4) for financial support.



\begin{thebibliography}{99}
\bibliographystyle{unsrt}

\bibitem{BKR} C.~Bourne, J.~Kellendonk, A.~Rennie, {\sl The $K$-Theoretic Bulk--Edge Correspondence for Topological Insulators}, Annales H. Poincar\'e {\bf 18},  1833-1866 (2017).

\bibitem{BP}
C.~Bourne, E.~Prodan, {\sl Non-commutative Chern numbers for generic aperiodic discrete systems}, preprint {\tt arXiv:1712.04136}.


\bibitem{BR} O.~Bratteli, D.~W.~Robinson, {\sl Operator Algebras and Quantum Statistical Mechanics 1}, (Springer, Berlin, 1979).

\bibitem{Con} A.~Connes, {\sl Noncommutative Geometry}, (Academic Press, San Diego, 1994).

\bibitem{EL} R.~Exel, T.~A.~Loring, {\sl Invariants of almost commuting unitaries}, J. Funct. Anal. {\bf 95}, 364-376 (1991).

\bibitem{FPL} 
I.~C.~Fulga, D.~Pikulin, T.~A.~Loring,
{\sl Aperiodic weak topological superconductors},
 Phys. Rev. Lett. {\bf 116}, 257002 (2016).

\bibitem{GVF} J.~M.~Gracia-Bond\'ia, J.~C.~V\'arilly, H.~Figueroa, {\sl Elements of noncommutative geometry}, (Springer Science \& Business Media, 2013).


\bibitem{GS} J.~Grossmann, H. Schulz-Baldes, {\sl Index pairings in presence of symmetries with applications to topological insulators}, Commun. Math. Phys. {\bf 343}, 477-513 (2016).


\bibitem{HL} M.~B.~Hastings, T.~A.~Loring, {\sl Topological insulators and C∗-algebras: Theory and numerical practice}, Annals of Physics {\bf 326}, 1699-1759 (2011).

\bibitem{Hir} M.~W.~Hirsch, {\sl Differential topology}, (Springer Science \& Business Media, 2012).

\bibitem{Lor} T.~A.~Loring, {\sl K-theory and pseudospectra for topological insulators}, Annals of Physics {\bf 356}, 383-416 (2015).

\bibitem{LS} T.~A.~Loring, H.~Schulz-Baldes, {\sl Finite volume calculations of $K$-theory invariants}, New York J. Math. {\bf 22}, 1111-1140 (2017).

\bibitem{SL} T.~A.~Loring, T.~Shulman, {\sl Noncommutative semialgebraic sets and associated lifting problems}, Trans. Amer. Math. Soc., {\bf 364}, 721-744 (2012). 

\bibitem{PLB} E.~Prodan, B.~Leung, J.~Bellissard, {\sl  The non-commutative $n$-th Chern number $(n\geq 0)$}, J. Phys. A: Math. Theor. {\bf 46}, 485202  (2013).

\bibitem{PSB} E.~Prodan, H.~Schulz-Baldes, {\sl  Bulk and boundary invariants for complex topological insulators: From $K$-theory to physics}, (Springer Int. Pub., Szwitzerland, 2016).

\bibitem{Sak} S.~Sakai, {\sl Operator algebras in dynamical systems}, (Cambridge University Press, Cambridge, 1991).

\end{thebibliography}
\end{document}